\documentclass[aps,prl,preprintnumbers,floats,twocolumn,superscriptaddress,nofootinbib,longbibliography]{revtex4-1}
\usepackage{graphicx}
\usepackage{dcolumn}
\usepackage{bm}
\usepackage[english]{babel}
\usepackage[latin1]{inputenc}
\usepackage{graphicx}
\usepackage{epstopdf}
\usepackage{amsfonts}
\usepackage{comment}
\usepackage{url}
\usepackage{color}
\usepackage{epsfig}
\usepackage{mathrsfs}
\usepackage{amsmath,amssymb}
\usepackage{subfigure}

\usepackage{framed}
\usepackage[svgnames]{xcolor}
\definecolor{shadecolor}{named}{LightGrey}

\begin{document}

\graphicspath{{figure/}}
\selectlanguage{english}

\newcommand{\lay}[1]{^{[#1]}}
\newcommand{\ud}{\,\mathrm{d}}

\def\Erdos{Erd\"os}

\title{Layer aggregation and reducibility of multilayer interconnected networks}

\author{M. De Domenico}
\affiliation{Departament d'Enginyeria Inform\'atica i Matem\'atiques, Universitat Rovira I Virgili, 43007 Tarragona, Spain}

\author{V. Nicosia}
\affiliation{School of Mathematical Sciences, Queen Mary University of London, London E1 4NS, United Kingdom}

\author{A. Arenas}
\affiliation{Departament d'Enginyeria Inform\'atica i Matem\'atiques, Universitat Rovira I Virgili, 43007 Tarragona, Spain}

\author{V. Latora}
\affiliation{School of Mathematical Sciences, Queen Mary University of London, London E1 4NS, United Kingdom}
\affiliation{Dipartimento di Fisica ed Astronomia, Università di Catania and INFN, I-95123 Catania, Italy}

\begin{abstract}
Many complex systems can be represented as networks composed by distinct
layers, interacting and depending on each others.  For example, in
biology, a good description of the full protein-protein interactome
requires, for some organisms, up to seven distinct network layers,
with thousands of protein-protein interactions each.
A fundamental open question is then how much information is really
necessary to accurately represent the structure of a multilayer
complex system, and if and when some of the layers can indeed be
aggregated. Here we introduce a method, based on information theory,
to reduce the number of layers in multilayer networks, while minimizing
information loss. We validate our approach on a set of synthetic
benchmarks, and prove its applicability to an extended data set of
protein-genetic interactions, showing cases where a strong reduction
is possible and cases where it is not.  Using this method we can
describe complex systems with an optimal trade--off between accuracy
and complexity.

\end{abstract}

\maketitle

Network science has shown that characterising the topology of a
complex system is fundamental when it comes to understanding its dynamical
properties~\cite{albert2002statistical,newman2003structure,boccaletti2006complex}.
However, in most cases, the basic units of real-world systems are 
connected by different types of interactions occurring at comparable 
time scales. For instance, this is the case of social systems, 
in which the same set of people might have political or financial
relationships~\cite{padgett1993robust}, or might be interacting using
different platforms like e-mail, Twitter, Facebook, phone calls,
etc.~\cite{krack1987,faust}. Similarly in biological systems, basic  
constituents such as proteins can have physical, co-localization, 
genetic or many other types of interactions. 
Recently, it has been shown that retaining the whole multi-dimensional 
information \cite{cardillo2013emergence} in the 
modeling of interdependent
\cite{buldyrev2010,gao2011networks} and multilayer
systems 
\cite{mucha2010community,dedomenico2013mathematical,nicosia2013growing,kivela2013multilayer} leads to new non-trivial structural
properties~\cite{battiston2014structural,nicosia2014measuring,dedomenico2013centrality}
and unexpected levels of dynamical
complexity~\cite{gomez2013diffusion,dedomenico2013random,granell2013dynamical,
  radicchi2013abrupt,nicosia2013nonlinear}. On the other hand, nothing
is known about the inverse problem, that is under which circumstances a
more compact aggregated representation of a system can give the same 
information of a fully multilayer representation.  
Inspired by quantum physics, where a similar question emerges to
quantify the distance between mixed quantum states
\cite{majtey2005jensen}, we propose a method to aggregate 
some of the layers of a multilayer system without a
sensible loss of information. Our procedure is based on the
application of information theory to graphs, and allows to construct a
reduced representation of a multilayer network which provides a good
trade--off between accuracy and compactness.

\section{Quantifying the information content of a multiplex network}

In quantum mechanics, there are \emph{pure states}, describing the
system by means of a single vector in the Hilbert space, and
\emph{mixed states}, arising from composite quantum systems described
by a statistical ensemble of pure states. The most general quantum
system can then be described by the so-called {\em density operator}
$\bm{\rho}$, a semidefinite positive matrix with eigenvalues summing
up to $1$, which encodes all the information about the statistical 
ensemble of pure states of the system \cite{dirac1981principles}.   
A widely adopted descriptor to measure the mixedness of a
quantum system is given by the Von Neumann entropy, the natural
extension of Shannon information entropy to quantum operators,
although other definitions, satisfying extensivity or non-extensivity
paradigms, have been lately introduced and studied
\cite{rossignoli2010generalized}. The Von Neumann entropy is defined
for any density operator $\bm{\rho}$. In particular, if the Von
Neumann entropy is zero, then the system is in a pure state, otherwise
it is in a mixed state.

The quantum mechanics formalism can also be used to describe a complex
system, if we imagine that each layer of a multilayer network
represents one possible state of the system, so that the entire
network is described by an ensemble of states.  It has already been
shown that the amount of information carried by a single-layer network
can be quantified by the Von Neumann entropy of the
graph~\cite{braunstein2006laplacian}, represented by a 
matrix which resembles quantum operators. 
Such a matrix can be obtained from the
Laplacian associated to the graph, after a proper normalisation (see
Appendix for details). The resulting number effectively summarises the
complexity of the wiring patterns of a network. Here we propose to use the Von
Neumann entropy to quantify the information gained or lost by
aggregating some of the layers of a multiplex network in a single
graph. This problem is surprisingly related to the separability of
mixing states in quantum systems
\cite{horodecki1996separability,rossignoli2002generalized,canosa2002generalized},
where, for instance, the entanglement of pure states is quantified by
considering the minimum information loss due to a complete local
measurement, in terms of the corresponding density operator
$\bm{\rho}$.

If the existence of inter-layer connections
  among the different replicas of the same node at the various layers of a multi-layer network 
is implicitly assumed, while their weights can not be defined or is not taken into account, the network is a multiplex network,  i.e. an edge-colored multi-graph \cite{kivela2013multilayer} that can be represented \cite{nicosia2013growing} by a set $\bm{A}=\{A\lay{1}, A\lay{2},
  \ldots, A\lay{M}\}$, whose elements are the adjacency matrices of the $M$ layers. 
  Placing each adjacency matrix of $\bm{A}$ in the diagonal of
  a $(N\times M)\times(N\times M)$ block matrix, while setting to zero the entries 
  in off-diagonal blocks, the above representation is casted into a 
  \textit{supra-adjacency matrix} \cite{gomez2013diffusion} $\mathcal{A}$, 
  a special flattening of the rank-4 adjacency tensor,   
  a more general representation of multilayer networks \cite{dedomenico2013mathematical}.
  Exploiting this mathematical representation, the Von Neumann
  entropy $h_{\mathcal{A}}$ of the interconnected multilayer network is computed
using Eq.~(\ref{eq:h_eigenvalues}) (see Appendix), as a function of the
$N\times M$ eigenvalues of the normalised Laplacian supra-matrix associated
to $\mathcal{A}$ \cite{dedomenico2013mathematical}. In the specific case of 
an edge-colored multi-graph this entropy reduces to the 
sum of the Von Neumann entropies of its layers, i.e., 
$h_{\bm{A}} = \sum\limits_{\alpha=1}^{M} h_{A\lay{\alpha}}$
where $h_{A\lay{\alpha}} = -\sum\limits_{i=1}^{N}
  \lambda_i\lay{\alpha} \log_2 (\lambda_i\lay{\alpha})$, and
  $\lambda_i\lay{\alpha}$ are the eigenvalues of the corresponding
  Laplacian matrix of $A\lay{\alpha}$.

\section{Quantifying the information loss in (partially) aggregated multiplex networks}

\begin{figure*}[!t]
 \includegraphics[width=18cm]{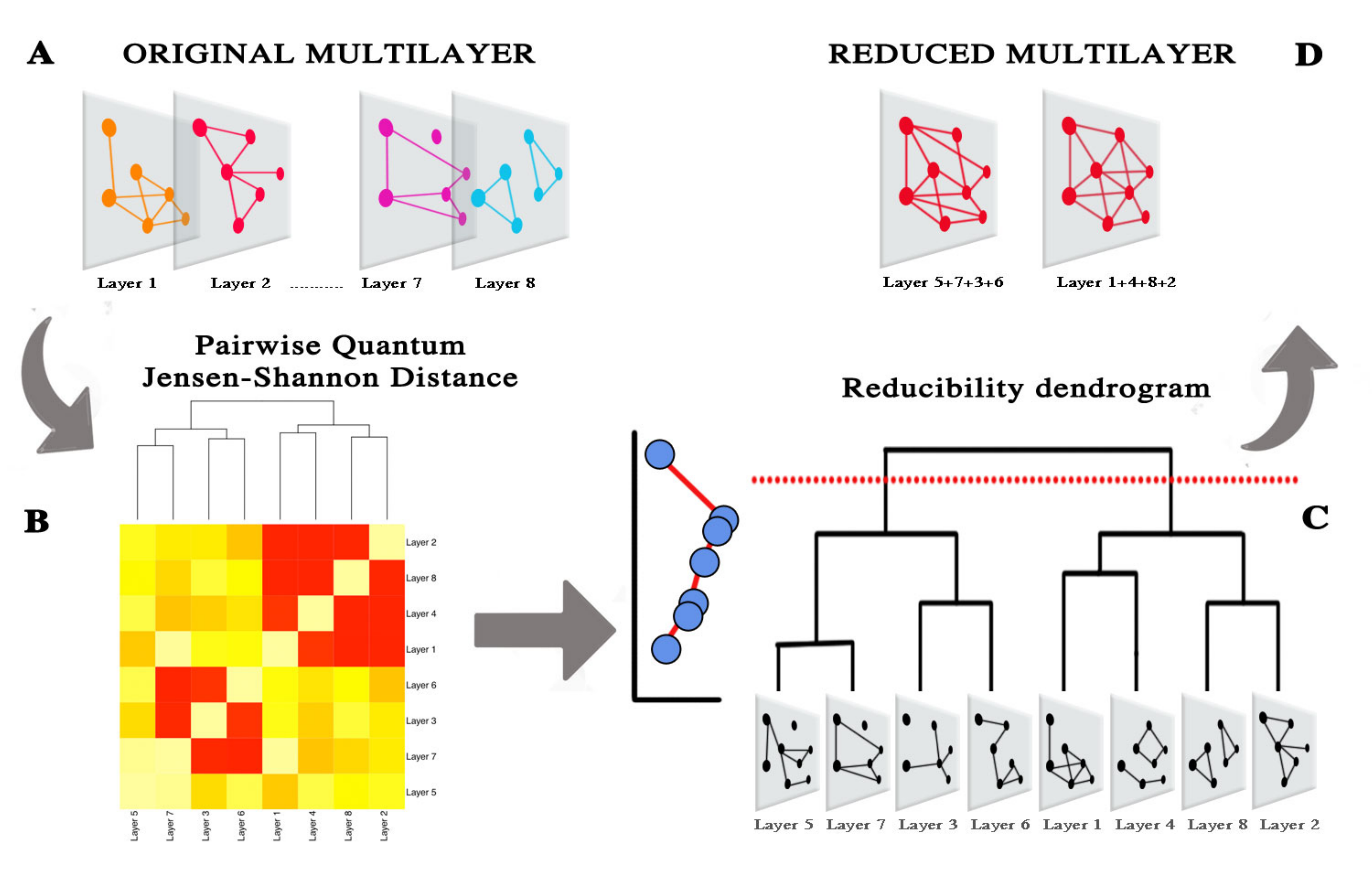}
 \caption{\textbf{Layer aggregation of multilayer networks.}
   Given a multiplex network \textbf{(A)}, we compute the the
   Jensen-Shannon distance between each pair of its layers
   \textbf{(B)}, which is a proxy for layer redundancy. The resulting
   distance matrix allows to perform a hierarchical clustering, whose
   output is a hierarchical diagram (a dendrogram) whose leaves
   represent the initial layers and internal nodes denote layer merging
   \textbf{(C)}. At each step, two layers (or group of layers) are
   merged and the information gain (or loss) is quantified by the
   global quality function $q(\bullet)$, shown by the curve on the
   left-hand side of panel \textbf{C}. The merging procedure is
   stopped when $q(M)$ is maximum, obtaining a reduced version of the
   original multiplex network \textbf{(D)}.}
 \label{fig:fig1}
\end{figure*}

Storing, handling and manipulating multiplex networks requires an
amount of space and computational power which increases at least
linearly with the number of layers of the system. It is therefore
natural to ask whether the additional information obtained by
explicitly considering the $M$ available layers of a system as
separate levels is indeed necessary to characterise it, or if instead
the dimensionality of the network can be reduced without a sensible
loss of information by aggregating some of the layers which carry
redundant information.

The Von Neumann entropy of a multiplex explicitly depends on the
actual number and structure of layers of which it consists, its value
being larger than the Von Neumann entropy of the corresponding
aggregated graph, by construction. To measure the information loss due
to the aggregation of a $M$-layer multiplex in a single-layer graph,
we use the relative entropy
\begin{equation}
  q(M) = 1 -
  \frac{1}{M} \frac{h_{\otimes}}{h_{\oplus}}
  \label{eq:relative_entr}
\end{equation}
where $\bigotimes= A\lay{1} \otimes A\lay{2} \otimes \ldots \otimes
A\lay{M}$ is the multiplex where all the $M$ layers are kept
separated, $\bigoplus = A\lay{1} \oplus A\lay{2} \oplus \ldots \oplus
A\lay{M}$ is the associated single-layer aggregated graph and
$h_{\otimes}$ and $h_{\oplus}$ are the corresponding Von Neumann
entropies. The rescaling factor $M^{-1}$ is necessary for a correct
comparison between $h_{\oplus}$ and $h_{\otimes}$.  The quantity
$q(M)$ measures the additional information obtained by considering a
$M$-layer multiplex representation of the system instead than a
single-layer aggregated graph. In particular, if all the layers of the
multiplex are identical then $q(M)=0$, since no layer adds new
information to that already encoded in the corresponding aggregated
graph. Conversely, higher values of $q(M)$ indicate that the $M$-layer
representation is more informative than a single-layer aggregation. We
notice that it is possible to obtain higher values of the relative
entropy in Eq.~(\ref{eq:relative_entr}) by considering a $\ell$-layer
multiplex where each layer corresponds either to one of the original
layers or to the aggregation of some of them.

In general, the optimal configuration of aggregated layers is the one
which maximises $q(\bullet)$, but finding such a configuration would
in general require the enumeration of all the possible partitions of a
set of $M$ objects (the layers), which is a well--known NP--hard
problem (i.e., its solution requires a computational time which scales
exponentially with $M$). To overcome this problem, we employ a
different approach, similar in spirit to the one adopted in quantum
physics to quantify the distance between mixed quantum states
\cite{majtey2005jensen}. More specifically, capitalizing on the concept
of Von Neumann entropy of a graph, we use the quantum Jensen--Shannon
divergence to quantify the (dis-)similarity between all pairs of
layers of a multiplex (see Eq.\,(\ref{eq:JSD}) and Appendix). This
choice is justified by the peculiar mathematical properties of this
measure, which allows to define a metric distance and can be used to
perform a hierarchical clustering of the layers. The result of this
procedure is a dendrogram (see Fig.~\ref{fig:fig1}), i.e., a
hierarchical diagram in which each of the $M$ leaves is associated to
one of the original layers of the system, each internal node indicates
the aggregation of (clusters of) layers into a single network and the
root corresponds to the fully aggregated graph. After the $m^{\rm th}$
step of the algorithm, we obtain a new multiplex network consisting of
$M-m$ layers, for which we can compute the associated value of
relative entropy $q(M-m)$. The cut of the dendrogram with maximal
value of $q(\bullet)$ corresponds to the (sub-)optimal configuration of
layers in terms of relative information gain with respect to the
aggregated graph.

The whole procedure proposed here is sketched in Fig.~\ref{fig:fig1}
and can be summarised as follows: i) compute the quantum
Jensen-Shannon distance matrix between all pairs of layers; ii)
perform hierarchical clustering of layers using such a distance matrix
and use the relative change of Von Neumann entropy as the quality
function for the resulting partition; iii) finally, choose the
partition which maximises the relative information gain.

\subsection{Layer aggregation of synthetic multiplex networks} 

\begin{figure*}
  \begin{center}
 \includegraphics[width=16cm]{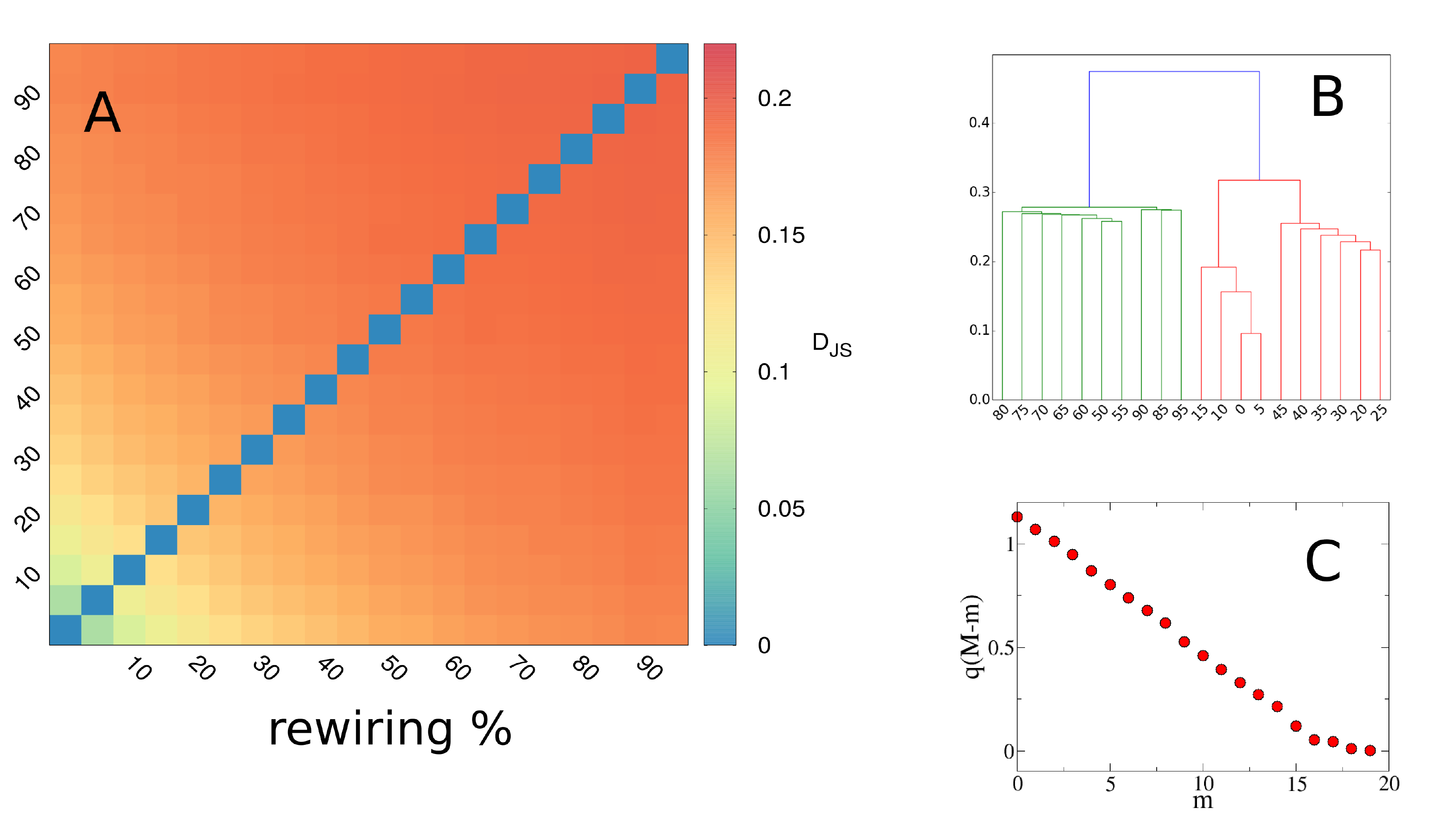}
  \end{center}
 \caption{\textbf{Multiplex benchmark}. We considered a benchmark
   multilayer network with $N=5000$ nodes and $M=20$ layers. The
   first layer is a scale-free graph with $P(k)\sim k^{-3}$, while the
   other layers are obtained by rewiring an increasing percentage of
   the edges of the first layer, from $5\%$ up to $95\%$. By doing so,
   each pair of layers is characterised by a different amount of edge
   redundancy (the total overlap of the multiplex is $<5\%$). (a) The
   heat map shows the Jensen-Shannon distance between the twenty
   layers, where each layer is identified by the corresponding
   percentage of rewiring. (b) The hierarchical clustering procedure
   successively merges layers with a decreasing percentage of
   redundant edges. (c) In this case $q(M-m)$ is a decreasing function
   of $m$, since each layer has some unique edges which are not
   present in the others. Consequently, the best representation of the
   multiplex is that in which all the layers are kept separated.}
 \label{fig:fig2}
\end{figure*}

To shed light on the impact of the layer aggregation procedure
proposed here on the structural properties of a multiplex network, we
considered different benchmarking scenarios. Each benchmark consists
of several layers characterised by specific features or a given amount
of correlation. In Fig.~\ref{fig:fig2} we report the case of a
multiplex network in which the layers are obtained by rewiring
different percentages of the edges of the first layer. The layers of
the resulting multiplex network are characterised by an increasing
amount of edge overlap. As shown in the Figure, the hierarchical
clustering procedure first aggregates layers which are more similar to
each other (namely, the layers which correspond to an amount of edge
rewiring smaller than $50\%$) and then merges the layers characterised
by higher rewiring. The monotonically decreasing behaviour of the
relative entropy $q(\bullet)$, shown in Fig.~\ref{fig:fig2}(c),
confirms that in this case the best representation of the system is
the one in which all the layers are kept distinct. In fact,
independently of the fraction of edges actually rewired, on average a
pair of layers exhibits a relatively small redundancy, since each of
the rewired layers carries some information which is not included in
the other layers (this multiplex has an overall overlap smaller than
$5\%$).

The results obtained from synthetic multiplex networks suggest that
layers with high overlap and similar topology tend to be aggregated
first. This corroborates our procedure showing that the principle of
minimum information loss is satisfied.

\subsection{Layer aggregation of multilayer biological networks}

To test the usefulness of our proposal on real-world networks, we
consider here the multiplex networks obtained by taking into account
different types of genetic interactions in 13 organisms of the
Biological General Repository for Interaction Datasets (BioGRID). This
is a public database that stores and disseminates genetic and protein
interaction information from model organisms and humans
(thebiogridd.org), and currently holds over 720,000 interactions
obtained from both high-throughput data sets and individual focused
studies, as derived from over 41,000 publications in the primary
literature. We use BioGRID 3.2.108 (updated to 1 Jan 2014)
\cite{stark2006biogrid}. In this data set, the networks represent
protein-protein interactions and the layers correspond to interactions
of different nature, i.e., physical (labelled ``Phys'' in the following), direct (``Dir''), co-localization (``Col''),
association (``Ass''), suppressive (``GSup''), additive (``GAdd'') or synthetic genetic (``GSyn'')
interaction. The number of layers identified for each organism ranges
from 3 to 7.

\begin{figure}
  \begin{center}
    \includegraphics[width=8cm]{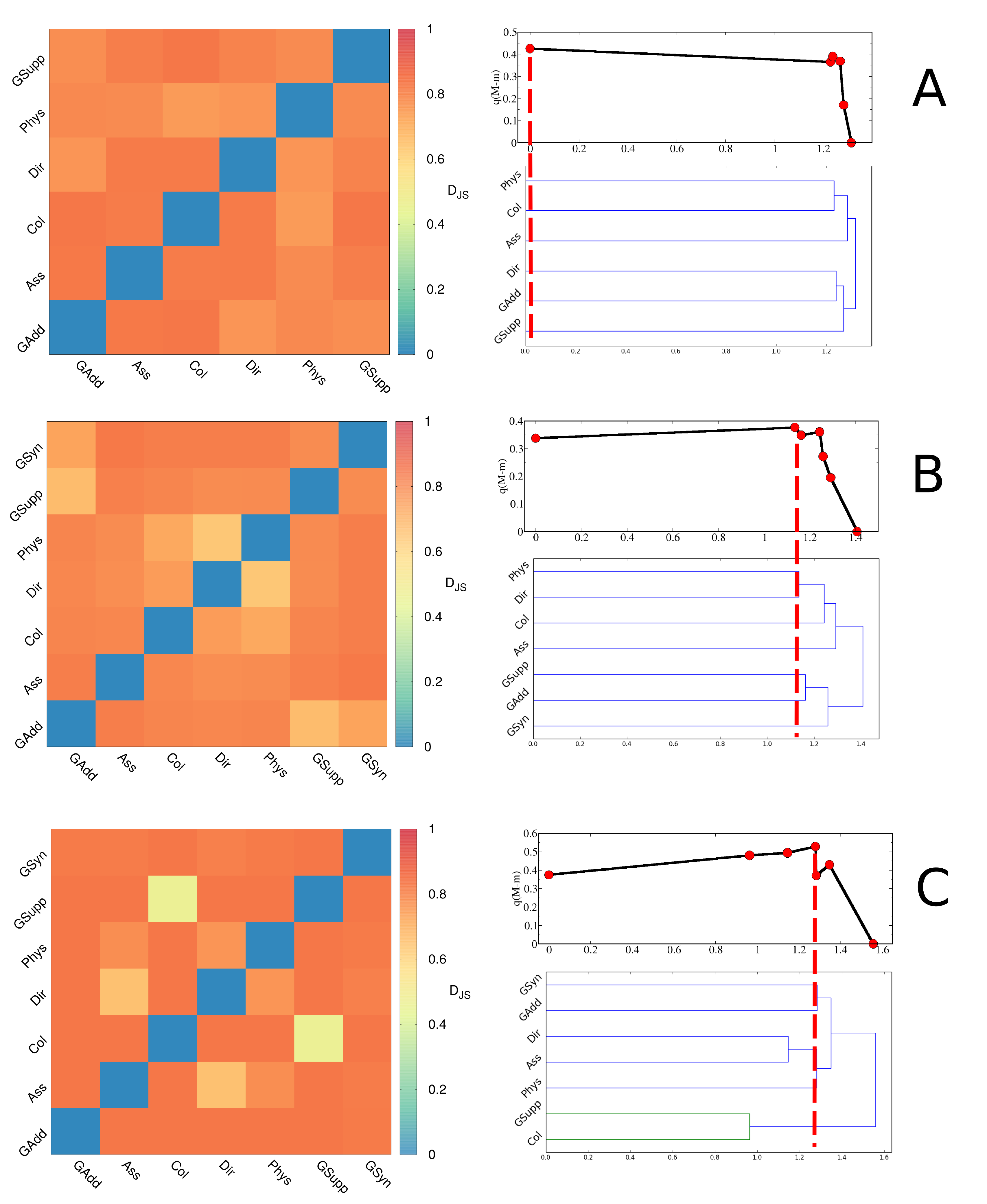}
  \end{center}
  \caption{\textbf{Layer aggregation of protein-genetic
      interaction networks.} The multiplex protein-genetic networks of
    different species have different levels of reducibility. We show
    the heat map of the Jensen-Shannon divergence, together with the
    dendrogram resulting from hierarchical clustering and the
    corresponding values of $q(\bullet)$, in three of the 13 species
    considered in this study. The dashed red lines identify the
    maximum of the global quality function $q(\bullet)$. For some
    organisms (like \emph{C. elegans}, reported in panel (A)), such maximum is
    obtained by leaving all the layers separate and no aggregation is
    possible, while for some other species a few layers carry
    redundant information, e.g. in (B) \emph{Mus} and in (C) \emph{Candida}, and can
    be safely compressed without sensible loss of information. }
  \label{fig:fig3}
\end{figure}

In Fig.~\ref{fig:fig3} we show the results obtained on three organisms
(\emph{C. elegans}, \emph{Mus} and \emph{Candida}). Despite the multiplex networks
corresponding to these organisms have a similar number of layers (six
for \emph{C. elegans}, seven for \emph{Mus} and \emph{Candida}), each of them is
characterised by a peculiar level of reducibility. In particular, in
the case of \emph{C. elegans} no layer aggregation is advisable at all, since
the maximum value of $q(\bullet)$ is obtained for the multiplex in
which all the six layers are kept distinct. Conversely, in the case of
\emph{Mus} and \emph{Candida} some pairs of layers carry redundant information and
can be aggregated without a sensible loss of information.

\begin{figure}
  \begin{center}
 \includegraphics[width=8cm]{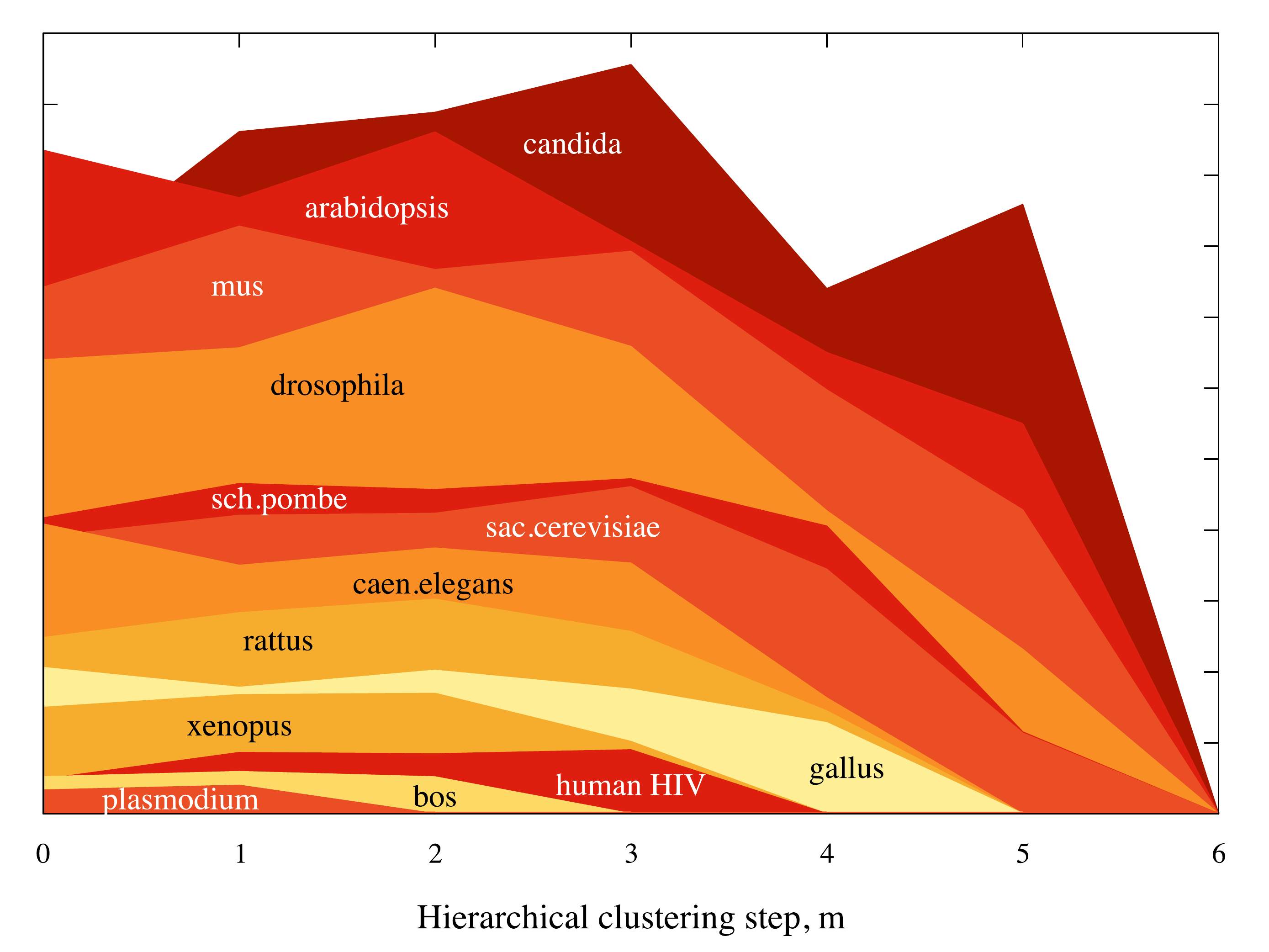}
  \end{center}
 \caption{\textbf{Reducibility of protein-genetic networks in the
     BioGRID data set}. The global quality function $q(\bullet)$ versus
   the number of merges in the hierarchical clustering procedure for
   the protein-genetic interaction multiplex networks of all the 13
   organisms considered in this study (the plots are vertically
   rescaled to avoid overlaps). The values of $q(\bullet)$ are not reported in the $y-$axis
   because only the existence of a global maximum, and the corresponding
   value of $m$ in the $x-$axis, is meaningful for the analysis.
   For each organism, $q(\bullet)$ has a
   maximum corresponding to the partition of the layers which
   minimises layer redundancy at the cost of a small loss of
   information.}
 \label{fig:fig4}
\end{figure}

 In Fig.~\ref{fig:fig4} we summarize the results obtained by applying
the proposed layer aggregation procedure to multilayer
genetic interaction networks in the BioGRID data set. This
particular visualization allows to compare the reducibility of all
organisms, simultaneously. Not all multiplex networks can be reduced
to a smaller number of layers, suggesting that for some organisms
layer aggregation should be avoided. For instance, this is
the case of \emph{m} (nematode),
\emph{Arabidopsis thaliana} (cress) and \emph{Bos taurus} (mammal),
where no global maximum is present -- except for $m=0$, i.e. the
initial multiplex. In other cases, reducibility might take place
as, for instance, in \emph{Saccharomyces cerevisiae} (yeast) and
\emph{Drosophila melanogaster} (common fruit fly), where a global maximum of
$q(\bullet)$ is present at $m=2$. 

We have proposed a practical procedure to aggregate layers of a multilayer network
and we have presented an application of our method to the case
of biological data sets. Nevertheless, our method is not limited to this 
kind of data and can be applied to other multilayer networks. For instance, 
we have applied it to the edge-colored multi-graph of European airports 
\cite{cardillo2013emergence}, where each layer indicates an airline, and we have
found that this transportation system can not be reduced to a smaller number of
layers. This result indicates that the connectivity between airports
is not redundant for any airline, as expected in the case of a large-scale 
transport infrastructure.

\begin{acknowledgments}
A.A. and M.D.D. are supported by MINECO through Grant
  FIS2012-38266; by the EC FET-Proactive Project PLEXMATH (grant
  317614) and the Generalitat de Catalunya 2009-SGR-838. A. A. also
  acknowledges partial financial support from the ICREA Academia and
  the James S. McDonnell Foundation. V.L. and V.N. aknowledge support
  from the Project LASAGNE, Contract No. 318132 (STREP), funded by the
  European Commission.
\end{acknowledgments}

\appendix

\section{References}
\vspace{-1truecm}
\bibliographystyle{aipauth4-1}
\bibliography{multiplex_reducibility}

\section{Appendix}

\subsection{Von Neumann entropy of single-layer networks} 
Given a graph $G(V,E)$ with $N=|V|$ nodes and $K=|E|$ edges,
represented by the adjacency matrix $A=\{a_{ij}\}$ where $a_{ij}=1$ if
node $i$ and node $j$ are connected through an edge, the Von Neumann
entropy of $G$ is defined as:
\begin{equation}
  h_G = -\text{Tr}\left[\mathcal{L}_{G} \log_2 \mathcal{L}_{G}\right]
  \label{eq:h_laplacian}
\end{equation}
where $L_{G}=I-D^{-1}A$ is the normalised Laplacian
associated to the graph $G$, $I$ is the identity matrix and
$D$ is the $N\times N$ diagonal matrix of node degrees
($d_{ij}=\delta_{ij}\sum_{\ell}a_{i\ell}$)~\cite{braunstein2006laplacian}. It
is easy to prove that $h_G$ can be written in terms of the set
$\{\lambda_1, \lambda_2, \ldots, \lambda_N\}$ of eigenvalues of
$L_G$:
\begin{equation}
  h_G = -\sum\limits_{i=1}^{N} \lambda_i \log_2 (\lambda_i).
  \label{eq:h_eigenvalues}
\end{equation}

\subsection{Jensen--Shannon distance between graphs} Given two density matrices $\bm{\rho}$ and $\bm{\sigma}$, it is
possible to define a (dis-)similarity between them by means of the
Kullback-Liebler divergence:
\begin{equation}
  \mathcal{D}_{KL}(\bm{\rho}||\bm{\sigma}) =
  \text{Tr}[\bm{\rho}(\log_2(\bm{\rho}) - \log_2(\bm{\sigma}))]
  \label{eq:KL}
\end{equation}
which represents the information gained about $\bm{\sigma}$ when the
expectation is based on $\bm{\rho}$ only. However,
$\mathcal{D}_{KL}(\cdot||\cdot)$ is not a metric, since it is not
symmetric with respect to its arguments (i.e.,
$\mathcal{D}_{KL}(\bm{\rho} || \bm{\sigma}) \neq
\mathcal{D}_{KL}(\bm{\sigma}||\bm{\rho})$) and it does not satisfy the
triangular inequality. A more suitable quantity to measure the
dissimilarity between two density operators is the Jensen--Shannon
divergence. If we call $\bm{\mu}=\frac{1}{2}(\bm{\rho} + \bm{\sigma})$
the new density matrix obtained as the mixture of the two operators,
the Jensen--Shannon divergence between $\bm{\rho}$ and $\bm{\sigma}$
is defined as:
\begin{eqnarray}
\label{eq:JSD}
\mathcal{D}_{\text{JS}}(\bm{\rho}||\bm{\sigma}) &=& \frac{1}{2}\mathcal{D}_{\text{KL}}(\bm{\rho}||\bm{\mu}) + \frac{1}{2}\mathcal{D}_{\text{KL}}(\bm{\sigma}||\bm{\mu}) \nonumber\\
&=& h(\bm{\mu}) - \frac{1}{2}[h(\bm{\rho})+h(\bm{\sigma})].
\end{eqnarray}
By definition, $\mathcal{D}_{JS}$ is symmetric and it is possible to
prove that $\sqrt{\mathcal{D}_{JS}}$, usually called Jensen--Shannon
distance, takes values in $[0,1]$ and satisfies all the properties of
a metric \cite{briet2009properties}.
Therefore, it can be used to quantify the distance, in terms of information gain/loss,
between the normalised Laplacian matrices associated to two distinct
networks.

\subsection{Hierarchical clustering}

Our idea is to measure the information lost by representing a
multiplex system through the single-layer network obtained by
aggregating all the available information in a single graph, with
respect to the case in which some (or all) of the constituent layers
of the system are kept distinct and separated. The main hypothesis is
that if the value of the Jensen--Shannon distance between the
Laplacian matrices associated to layers $\alpha$ and $\beta$ is small,
then the two layers can be safely merged in a single network without
loosing too much information. Conversely, if
$\mathcal{D}_{JS}(\mathcal{L}_{\alpha}, \mathcal{L}_{\beta})$ is
large, then the two layers provide different information about the
relationships among the nodes of the system. In this case, it would be
better to leave the two layers separated, since their aggregation will
result in a substantial loss of information.

We employ a classical hierarchical clustering of the $M$ layers of a
multiplex based on the Jensen--Shannon distance between layers. At
each step of the algorithm, we aggregate the two clusters of layers
which are separated by the smallest value of $\mathcal{D}_{JS}$, and
then we update the distances between the newly formed cluster and the
remaining ones according to Ward's linkage. By iterating this procedure
$M-1$ times we obtain a dendrogram, i.e. a hierarchical diagram whose
$M$ leaves are associated to the original layers of the system,
internal nodes indicate merges of (clusters of) layers and the root
corresponds to the aggregated graph.

\end{document}